\documentclass[aps,pre,groupedaddress,twocolumn]{revtex4-1} 
\usepackage{graphicx}
\usepackage[mathscr]{euscript}
\begin{document}
\title{The effect of inertia on sheared disordered solids: \\
Critical scaling
of avalanches in two and three dimensions }

\author{K. Michael Salerno} \affiliation{Department of Physics and
Astronomy, Johns Hopkins University, Baltimore, MD 21218, USA}
\author{Mark O. Robbins} \affiliation{Department of Physics and
Astronomy, Johns Hopkins University, Baltimore, MD 21218, USA}
\thanks{}
\date{\today} 
\begin{abstract} 

Molecular dynamics simulations with varying damping are used to examine the
effects of inertia and spatial dimension on sheared disordered solids in
the athermal, quasistatic limit.
In all cases the distribution of avalanche
sizes follows a power law over at least three orders of magnitude in dissipated
energy or stress drop. Scaling exponents are determined using finite-size
scaling for systems with ~$10^3$ to $10^6$ particles. Three distinct
universality classes are identified corresponding to overdamped and
underdamped limits, as well as a crossover damping that separates the two
regimes. 
For each universality class, the exponent describing the avalanche
distributions is the same in two and three dimensions.
The spatial extent of plastic damage is proportional
to the energy dissipated in an avalanche.  Both rise much more rapidly with
system size in the underdamped limit where inertia is important. Inertia
also lowers the mean energy of configurations sampled by the system and
leads to an excess of large events like that seen in earthquake
distributions for individual faults.
The distribution of stress values
during shear narrows to zero with increasing system size and may provide
useful information about the size of elemental events in experimental
systems. For overdamped and crossover systems the stress variation scales
inversely with the square root of the system size.  For underdamped systems
the variation is determined by the size of the largest events.

\end{abstract}
\pacs{}
\maketitle
\section{\label{Intro}Introduction}

Many slowly driven systems respond to a driving force through bursts of
activity termed avalanches.
These avalanches often follow a power-law distribution over many decades,
signalling the existence 
of a nonequilibrium critical depinning transition at the onset of motion. 
Systems showing this scaling behavior include charge-density waves, fluid interfaces, magnetic domain walls,
granular media, foams, crystals, amorphous metals and the earth's crust \cite{Martys1991PRB, Perkovic1995,Myers1993a, Miguel2001, Friedman2012,Tsekenis2010,Baldassarri2006,Zapperi1997, Zapperi2005, Sethna2001}.

One of the unresolved aspects of the depinning transition is the role of inertia.
Most computational work has focussed on the overdamped limit, and studies that include inertia
suggest that it fundamentally changes the depinning transition from a continuous
second order transition to a discontinuous transition with hysteresis
\cite{Dahmen2009,Schwarz2003,Prado1992,Marchetti2005}.
In contrast, experiments have reported power law scaling even in systems that display
underdamped dynamics, such as earthquakes and compressed laboratory samples.
In this paper we present simulations of quasistatic plastic deformation in two and three dimensional
disordered solids that show inertia does not destroy critical behavior at depinning, but does
change the universality class.

Theoretical studies of avalanches have generally considered lattice-based models
with simple site interaction rules
\cite{Alava2006,Cizeau1997,Dahmen2009,Friedman2012,Dahmen2011,Zapperi1997,
Prado1992,Kadanoff1989,Martys1991PRB,Maimon2004,Talamali2011,Zaiser2007,Bak1987, Picallo2008}.
These models are computationally and analytically tractable, but have the
limitation that position and stress changes are discrete. Most studies consider
scalar order parameters and are in the overdamped, mean-field limit where
correlations in deformation are ignored.  Then the rate of avalanches of size
$S$ follows a power law $R(S)\sim S^{-\tau},$ with a universal value of
$\tau=3/2$ \cite{Alava2006,Dahmen2011}.
Strain and stress in deformed solids are tensoral quantities rather than
scalars.  A recent model shows that this can produce a lower value of
$\tau=1.25$ \cite{Talamali2011} and produce longe-range spatial
correlations in deformation like those found in atomistic simulations
\cite{Maloney2009}.
Older models with different rules for site evolution, such as long-term damage
to sites, also find that spatial correlations affect the power-law exponents
for avalanche statistics \cite{Zapperi1997}.

Another limitation of lattice models is the difficulty of including inertia.
One common approach is to add rules that lower barriers to motion when
an avalanche starts \cite{Prado1992,Maimon2004,Dahmen2009,Friedman2012, Dahmen2011}.
This fits the intuitive picture that
inertia can carry a system over successive potential energy barriers, but inertia is
highly directional, and decreases the chance of passing over barriers that are
not in the direction of the momentum.
Lattice models of this type
find that inertia fundamentally changes the nature of the
depinning transition.  All find that inertia introduces hysteresis, with different
stresses required to initiate and stop motion
\cite{Maimon2004,Schwarz2001,Dahmen2009,Prado1992}.  Most predict the transition
becomes first order\cite{Schwarz2001,Dahmen2009,Prado1992}, but a hysteretic second order has also been proposed
\cite{Maimon2004}.
Experimental evidence for hysteresis and a first order depinning transition has been reported for granular media and sandpiles
\cite{Held1990,Jaeger1989}.

Many other expermients have reported a continuous second order transition
with critical scaling in plastically deformed
disordered solids, including granular packings,
colloidal glasses, foams and metallic glasses 
\cite{Schall2007,Utter2004, Utter2008,Hayman2011,
Miller1996,Wu2008894,PhysRevB.64.180201,Greer2011858,greer2010transition,Bretz2006}.
The longest range of scaling is for earthquakes, where the conversion
of the magnitude on the Richter scale to the energy or moment is complicated,
but gives $\tau=5/3$.\cite{scholz2002mechanics}
Studies of granular media have found values of $\tau$ that
are $2$ to $6$ \cite{Bretz2006,Hayman2011}, or as small as $1.2$
\cite{Baldassarri2006,Bretz2006,behringerEmail2013}.  
Sun and coworkers have reported results from deformation of bulk metallic glasses,
finding avalanche distribution exponents $\tau$ between $1.3$ and $1.5$
\cite{Sun2010,Sun2012}.
The variation in laboratory measurements reflects the difficulty in
detecting events with a wide range of sizes.
In addition, it is difficult to
vary the rate, system size, and other experimental parameters
that cut off the largest events and influence the apparent exponent \cite{Dahmen2012-tuned}.
Simulations allow full analysis of these effects.
 
Particle-based simulations of plastic deformation provide more realistic
microscopic detail than lattice models but are also computationally more intensive.
Early simulations of bubbles in the overdamped
limit found a power-law distribution of rearrangements with exponent
$\tau = 0.7$ \cite{Tewari1999,Durian1997}.
Maloney and Lema\^{\i}tre found similar scaling using energy-minimization
dynamics for quasi-static shear of a model two-dimensonal glass, but $\tau$
appeared to decrease from 0.7 to 0.5 with increasing system size
\cite{Maloney2004a}.
They also found that the size of the largest avalanche increased with system
size, as expected at a critical depinning transition.  A power law increase in
event size was seen in later simulations of a similar two-dimensional model by
Lerner and Procaccia \cite{Lerner2009} and three diemensional simulations of a
more realistic model of amorphous metals by Bailey et al. \cite{Bailey2007}.
The computational studies described above have all been on systems of about $2
\cdot 10^4$ particles or less, restricting the range of power law scaling, and
making $\tau$ difficult to measure.  

In a recent paper, we examined scaling of avalanches in two-dimensional systems
with more than a million particles.
This allowed $\tau$ and other critical
exponents to be determined as a function of damping.
The results showed that the depinning transition remained second order in
the underdamped limit, but that the universality class changed with damping.
The current paper expands our studies of two dimenional systems and extends them
to three dimensions.
All simulations are performed in the
athermal quasi-static limit, but with varying levels of damping to change the
role of inertia in the system.  Finite-size scaling is used to develop
scaling relations between critical exponents and analyze data from systems with
thousands to millions of particles. 

We find that inertia leads to the same three universality classes in two and
three dimensions.
Some of the exponents, including $\tau$, are independent of dimension.
 In the overdamped limit, $\tau$ is
close to the value of 1.25 obtained in a recent lattice model that includes
directional stress transfer \cite{Talamali2011}, and less than the value of 1.5
obtained for scalar lattice models
\cite{Dahmen2012-tuned,Dahmen2011,Dahmen2009}.
In the underdamped limit, $\tau$ is also near 1.25, but the distribution
of stress drops can exhibit a higher apparent exponent.
The most dramatic effect of inertia is to increase
prevalence of large avalanches.  The magnitude of the largest events grows more
rapidly with system size and the avalanche distribution has a plateau at large
magnitudes.  The longest range of scaling is observed for the crossover damping
that separates these two regimes.  The energy scales with $\tau=1$ for more than 5 decades.

There is no evidence of the hysteresis predicted by lattice models of
inertia \cite{Dahmen2009,Dahmen2011,Schwarz2003,Maimon2004,Schwarz2001,Marchetti2005}. 
The range of stresses sampled during quasistatic motion
shrinks to zero with increasing system size.  In the overdamped limit, stress
fluctuations decrease as the inverse square root of the number of particles.
Stress fluctuations drop more slowly in the underdamped limit, where they are
determined by the size of the largest avalanches.  In all cases, the rise in
maximum avalanche size with system size reduces the rate of small avalanches,
which scales sublinearly with system size.
Both effects may be useful in analyzing the effective size of elemental deformations in
experimental systems.

The relationship between the change in shear stress,
energy dissipation and plastic
deformation during individual avalanches is studied in detail.
The drop in stress and the energy dissipated
are uncorrelated
for small events, but become linearly related for the larger events that exhibit critical scaling.
The spatial size of
avalanches is proportional to the energy dissipated.
This provides evidence that bigger avalanches spread over a larger area rather than producing greater
local strains in the deformed region.

An outline of the remainder of the paper is as follows: The second section
of the paper details the system studied, including the particle
interactions, damping, and the protocol for reaching the quasi-static limit
in simulations with inertia.  The third section of the paper
presents results for the effect of inertia on the energies and stresses sampled
during shear, the rate of avalanches, critical scaling exponents, and the relation between the energy dissipated,
stress change and plastic deformation associated with each avalanche.
Finally, the fourth section
presents a discussion and summary of the findings.


\section{\label{Methods}Methods}

This paper presents results from molecular dynamics (MD)
simulations of deformed disordered solids in two and three dimensions. In
all cases, the system studied is a binary glass. The two species of
particles A and B both have mass $m,$ but have different diameters to
prevent crystallization.   The particles interact via a smoothed
Lennard-Jones (LJ) potential, which depends only on the magnitude $r$ of
the vector $\mathbf{r}$ between two particles and their species.  This
potential keeps the standard LJ form at small distances: 
\begin{equation}
U(\mathbf{r}) = 4 u [(a_{ij}/r)^{12} - (a_{ij}/r)^{6}] + u_c ~~ , ~~
r < 1.2a_{ij},
\end{equation} 
where $u$ is the characteristic energy, $u_c$ is
an energy offset, and $a_{ij}$ is the interaction length between particles
of type I and J. The $A-A$ particle interaction length is taken as the
fundamental unit of length $a \equiv a_{AA}.$  The $B-B$ particle
interaction length $a_{BB} = 3/5 a,$ while the mixed interaction length
$a_{AB} = a_{BA} = 4/5 a.$  Outside the LJ region the potential has a
polynomial form 
\begin{widetext}
\begin{equation} U(\mathbf{r}) = C_1(r-r^{(c)}_{ij}) +
\frac{C_2}{2}(r-r^{(c)}_{ij})^2 + \frac{C_3}{3}(r-r^{(c)}_{ij})^3 +
\frac{C_4}{4}(r-r^{(c)}_{ij})^4 ~~, ~~ 1.2a_{ij} < r < 1.5a_{ij},
\end{equation} 
\end{widetext}
with coefficients $C_i$ chosen so that the energy, force and
the derivative of the force match the LJ form at the inner cutoff radius, $
1.2a_{ij} $, and go to zero at an outer cutoff radius, $ r^{(c)}_{ij} = 1.5
a_{ij} $.
 For these cutoff radii the binding energy of a single
bond is about 0.5 $u$.
 The strength of the interaction, particle radius, and mass  set
the fundamental unit of time, $t_0 = \sqrt{m a^2/u}.$  Simulations were
performed with the LAMMPS MD simulation code, using a velocity Verlet
integration algorithm with an integration timestep $\Delta t = t_0/200$
\cite{Plimpton1995}.  

Two dimensional systems are initialized by placing particles at random in a
square periodic simulation cell with the ratio of the number of particles
of species A and B: $N_A/N_B = (1+\sqrt{5})/4$.  Next, the system is heated
well above the glass transition temperature and then quenched to zero
temperature at constant pressure.  The pressure is chosen to be near zero
but slightly compressive.  The results are insensitive to the precise
value.  Following this procedure, the system density is
$\rho = 1.38a^{-2} $ and the square simulation box has period L.  We
consider five box sizes with $L = 55a, 109a, 219a, 437a $ and $ 875a$.
These sizes correspond to $N \approx 4 \cdot 10^3 $ to $10^6 $ particles.

A similar equilibration protocol is followed for three dimensions.  After
equilibration the density is $ \rho = 1.71a^{-3} $ and the simulation is a
cube with period L.  Sizes are $L = 20a, 40a,$ and $81a$,
corresponding to $N \approx 10^4 $ to $10^6$ particles.  

After the quench process, the samples are strained.  The periodic boundary
conditions are changed, and a corresponding affine displacement is applied
to the position of each particle.   The deformation applied to the
simulation box in two dimensions is a pure shear strain at a true strain
rate $ \dot{\epsilon} = \dot{\epsilon}_{xx} = -\dot{\epsilon}_{yy} $. In
three dimensions the system volume is conserved by applying an axisymmetric
compressive strain-rate $-\dot{\epsilon} $ in two dimensions (x and y) and
an extensional strain rate $2\dot{\epsilon}$ in the third direction (z).  

\begin{figure}[placement ] 
\includegraphics[width=0.35\textwidth]
{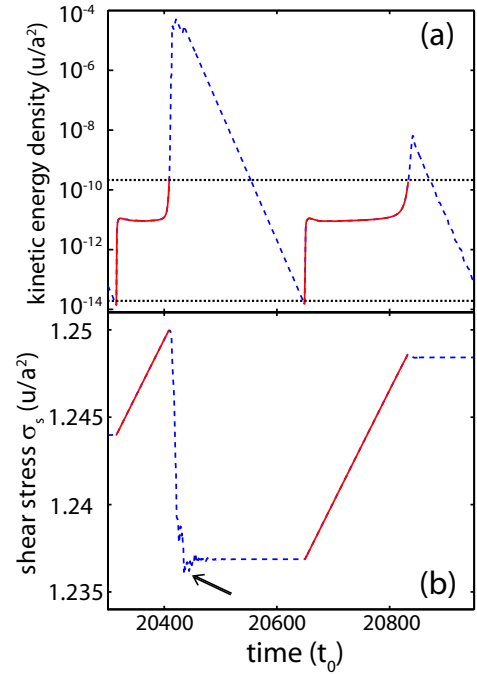}
 \caption{\label{quasi-example} (color online) (a)
An event sequence showing the kinetic energy thresholds (dotted horizontal
lines) used to reach the quasi-static limit.  Here an $L=219a$ system is
strained at a rate of $\dot{\epsilon} = 10^{-6}t_0^{-1}$ during the solid
(red) segments and the low, constant kinetic energy reflects non-affine
displacements due to heterogeneity.  Avalanches cause a sharp spike in
kinetic energy that decays more rapidly as $\Gamma$ increases.  The strain
rate is set to zero (dashed blue segments) after the upper threshold is
exceeded and returned to $10^{-6}t_0^{-1}$ when a lower threshold is
passed.
(b) The stress-strain curve rises linearly during elastic loading (solid
red) and drops rapidly as the avalanche begins (dashed blue).  There is
often an overshoot (arrow) where the stress drops below the steady-state
value.} \end{figure}

Our aim is to study the athermal limit, which requires constantly removing
kinetic energy from the simulation.  A viscous drag force is applied to
damp particle motion.  The drag force has the form $\mathbf{F}_{drag} =
-\Gamma m \mathbf{v}$ where $\mathbf{v}$ is the peculiar velocity, with
displacement due to the affine deformation subtracted.  The dissipation
rate $\Gamma$ plays a central role in our simulations by controlling the
relative importance of the inertial term in the particle equations of
motion.  As $\Gamma$ decreases, the dynamics changes from overdamped to
underdamped (inertial) dynamics.
  
Generically, a strained disordered solid will load elastically for some
strain interval and then plastically deform, decreasing the stress in the
system and releasing stored elastic energy as kinetic energy.  These sudden
bursts of particle motion are termed avalanches.  In the quasi-static limit
the series of elastic loading segments and plastic deforming avalanche
events should be independent of strain rate and depend only on the total
strain interval.  One way to realize this limit is to deform the system at
a very low rate, so that the kinetic energy from one avalanche has been
dissipated long before the system has been strained enough to nucleate the
next avalanche. Since the rate must be set low enough to prevent
overlap of the closest events, this is not computationally feasible for all
system sizes and damping rates.  Instead, we implement a protocol where the
system is strained at a finite strain rate, which is then reduced to zero
when an avalanche is detected.

A representative strain-avalanche-strain interval, shown in
Fig.\ref{quasi-example}, illustrates how the system evolves with this
quasi-static avalanche detection scheme.  When the system is deformed, the
non-affine response due to heterogeneity in the solid produces a small
background kinetic energy density, $KE_{back}$.  This kinetic energy is
nearly constant during elastic loading at constant strain rate (solid
lines).  When an avalanche starts, there is a sharp rise in kinetic energy.
The strain rate is reduced to zero when the kinetic energy exceeds
$KE_{back}$ by roughly two orders of magnitude.  The straining of the solid
resumes when the kinetic energy has fallen below $KE_{back}$ by at least
two orders of magnitude.  We have checked that the strain rate chosen is
low enough that the results are not sensitive to these thresholds.  

The stress response of the system illustrated in Fig. \ref{quasi-example}b
is typical, showing linear behavior during the strain interval, followed by
a rapid drop during an avalanche event.  During avalanches with low
particle damping the stress often overshoots the steady-state value which
is used to quantify the size of avalanches.  This overshoot complicates
analysis of simulations at constant strain rate.  

Even after eliminating the connection between strain rate and avalanche
duration, there is still the problem of the very long duration of large
avalanches at very low damping rates.  As the damping coefficient $\Gamma $
becomes small and events become large, the peak kinetic energy in the
system approaches $10^{-3} u$ per unit area (volume).  At our prescribed
strain rates there is then a factor of $ KE_{max}/KE_{back} \approx 10^{8}$
between the maximum kinetic energy and the kinetic energy during straining.
This energy must be removed by the viscous drag force and one can estimate
that for the smallest damping rates we simulate, $\Gamma t_0 = 10^{-3} $,
the decay of the kinetic energy will take a time of about $
\log(KE_{max}/KE_{back})/\Gamma \approx 20,000 t_0$.  This is not only
computationally expensive, but unnecessary.  Even for our largest system
sizes the time for sound waves to propagate across the system, $ t_{prop} =
L/c \approx 250 t_0, $ is much smaller.  
 Underdamped systems seldom show signs of
further instability, such as kinetic energy spikes or stress drops, after
about $2-3 t_{prop}$.  

In order to expedite draining the system of kinetic energy when $\Gamma t_0
= 10^{-3},$ we quench the kinetic energy rapidly once a threshold  has been
reached.  Our criterion is that when the kinetic energy in the system has
fallen to about $10^{-3}$ times the peak kinetic energy the avalanche is
effectively over and no other instabilities will be activated.   For the
viscous damping force discussed above, this is equivalent to a time
criterion since the decay of the kinetic energy in the system is
exponential. For comparison, the time for this decrease in kinetic energy
is still roughly ten times larger than the timescale for sound to propagate
across the largest simulation cell ($L=875a$).  It is also much larger
than the time for the stress to reach its steady-state value (Fig.
\ref{quasi-example}), indicating an event is over.  

In order to verify that the quench procedure does not affect system
evolution, we compared it to simulations with constant damping.  For a
subset of avalanches simulated with both the "quench" protocol and fixed
damping rate, the total energy dissipated differed by less than $ 10^{-9}
u$.  This is orders of magnitude smaller than the smallest avalanches
recorded, which have energy $E \approx 10^{-5} u$.  We conclude that the
quench protocol produces a system in the same local potential energy
minimum as the unquenched simulation. 

\begin{figure}[placement h!]
\includegraphics[width=0.45\textwidth]
{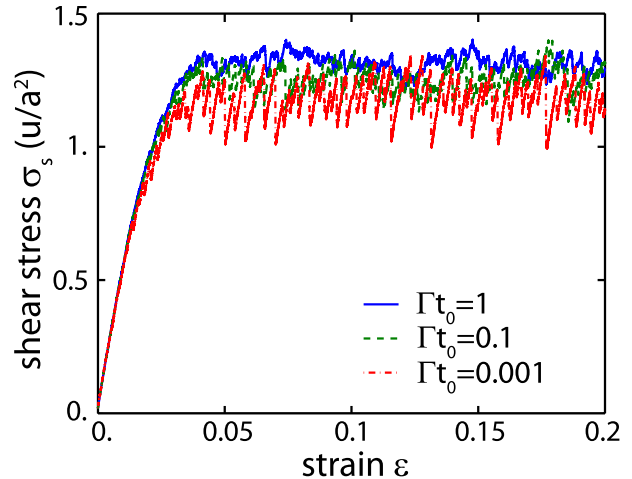}
\caption{\label{stress_strain}(color online) Typical stress-strain curves
in two dimensions for system size $L=219a $ and three different damping
rates:  $\Gamma t_0 = 1$ (solid blue), $\Gamma t_0 = 0.1$ (dashed green),
and $\Gamma t_0 = 0.001$ (dash-dot red). } \end{figure}

Typical two-dimensional stress-strain loading curves are shown in 
Fig. \ref{stress_strain}.  Curves for three dimensional systems are similar.
For all damping rates the systems reach steady-state after roughly 5\%
strain.  There is a small drift in the hydrostatic pressure at larger
strains, but quantities of interest like the shear stress and avalanche
statistics become stationary and do not evolve with strain.   Only
avalanches at strains greater than 7\% are included below.
 
Generically, the elastic energy density stored in a system by a
differential strain $d\epsilon$ is  $ du_{strain} =  \sigma_{ij}
d\epsilon_{ij},$  where $\sigma_{ij} $ is the stress tensor, and summation
over repeated indices is implied.  Because the 2D strain geometry is pure
shear, this can be simplified by defining $ \epsilon \equiv \epsilon_{xx} =
-\epsilon_{yy} $ and $\sigma_s \equiv \sigma_{xx} - \sigma_{yy} $.  The
stored elastic energy density is then  \begin{equation}\label{simpstress}
du_{strain} = \sigma_s d\epsilon . \end{equation}  The elastic strain
energy in three dimensions has the same form if one defines $\sigma_s
\equiv \sigma_{xx} + \sigma_{yy} - 2\sigma_{zz} $ and $\epsilon \equiv
\epsilon_{xx} = \epsilon_{yy} = -1/2\epsilon_{zz} $.

\section{\label{results}Results}

\subsection{\label{Timedep}Time Dependence of Stress and Energy}

\begin{figure}[placement h!]
\includegraphics[width=0.45\textwidth ]
{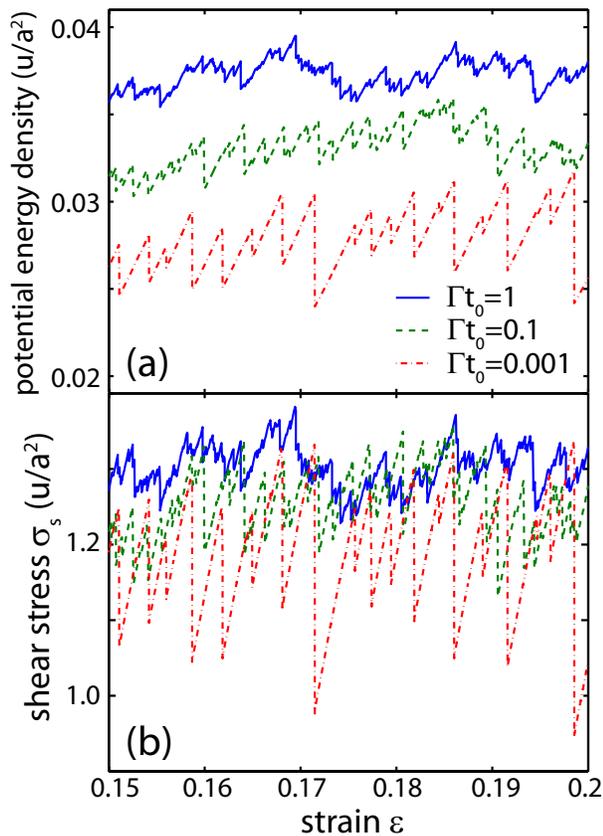}
\caption{\label{stress-pe-strain} 
(color online) The (a) potential energy and (b) stress in three systems
during a representative  strain increment.  The three systems of length $L
= 219a$ started at the same zero-strain particle configuration but evolved
with different, representative damping rates: $\Gamma t_0 = 1$ (solid
blue), $\Gamma t_0 = 0.1$ (dashed green) and $\Gamma t_0 = 0.001$ (dash-dot
red). The mean values and the size and rate of fluctuations in energy and
stress vary significantly with damping rate. }
\end{figure}

This section illustrates some of the dramatic effects that inertia has on
the mean of and fluctuations in the shear stress and potential energy
density in steady-state, quasi-static shear.  One limiting case is the
overdamped (large $\Gamma$) regime, where the potential energy decreases
monotonically to the next minimum during each avalanche.  In the opposite,
underdamped limit, there is negligible damping during plastic
rearrangement, and inertia can carry the system over successive small
energy barriers.  We present typical results from these limiting regimes
with damping rates  $\Gamma t_0 = 1$ and $\Gamma t_0 = 0.001$,
respectively.  Ref. \cite{Salerno2012} identified a critical intermediate
damping rate of $\Gamma t_0 = 0.1$ that separates these regimes.  We find
that this crossover damping rate is the same in 2D and 3D within our
uncertainty.

Figure \ref{stress-pe-strain} illustrates how damping affects the potential energy
density and stress.  Note that systems with different damping rates sample
completely different potential energy landscapes with almost no overlap.
Differences of approximately 30\% in the mean value of the potential energy
density persist in our largest system sizes in two dimensions.  There is a
smaller but significant change of about 10\% in the mean stress.  In three
dimensions the mean potential energy in the overdamped and underdamped limits
varies by about 8\% and the stress by 6\%.
  As the damping decreases, inertia is able to carry
the system over barriers in the potential energy landscape to progressively
lower minima.  In addition to reducing the mean potential energy, inertia leads
to larger avalanches.  The increase in the size of energy and stress drops is
evident in Fig.  \ref{stress-pe-strain} and related to changes in scaling
exponents discussed below.

As illustrated in Fig. \ref{stress-pe-strain}, the evolution of the stress and
potential energy density is characterized by linear rises, where elastic
energy is stored, and sudden drops during avalanches.  Each avalanche can
be characterized by the potential energy density drop $\Delta \mathcal{U}$ and stress
drop  $\Delta \sigma_s.$  In what follows we want to compare avalanches of
the same absolute size in systems of different linear dimension $L$.  We
define absolute measures of energy and stress drop as
\begin{equation} E
\equiv L^d\Delta \mathcal{U} ,\quad S \equiv \frac{\langle \sigma_s\rangle L^d}{4\mu}
\Delta \sigma_s \qquad.
\end{equation} 
The shear modulus $\mu$ and the
steady-state shear stress $\langle \sigma_s\rangle $ are introduced so that
both $S$ and $E$ have units of energy.   We have found that both $\mu$ and
$\langle \sigma_s\rangle $ are nearly independent of system size and
relatively insensitive to damping rate. 

With these normalizations, energy conservation imposes a sum rule:
\begin{equation}\label{sum_rule_eq} \sum_i E_{i} = \sum_i S_{i},
\end{equation} where the sum is over all avalanches $i$ in the steady-state
regime.  This connection has been noted previously by Lerner and Procaccia
\cite{Lerner2009} and is derived in Appendix \ref{sum_rule}.

The sum rule in Eq. \ref{sum_rule_eq} only constrains the sum over all
events, but one might expect that something similar to the principle of
detailed balance leads to a correlation between $E$ and $S$ for individual
events.  This correlation clearly breaks down for small events.  Indeed,
while E is always positive, $S$ can have either sign for small events
\cite{Lerner2009}.  Large events dominate the sums in Eq.
\ref{sum_rule_eq} and their energy and stress drops are more strongly
correlated.

\begin{figure}[placement ]
\includegraphics[width=0.45\textwidth]
{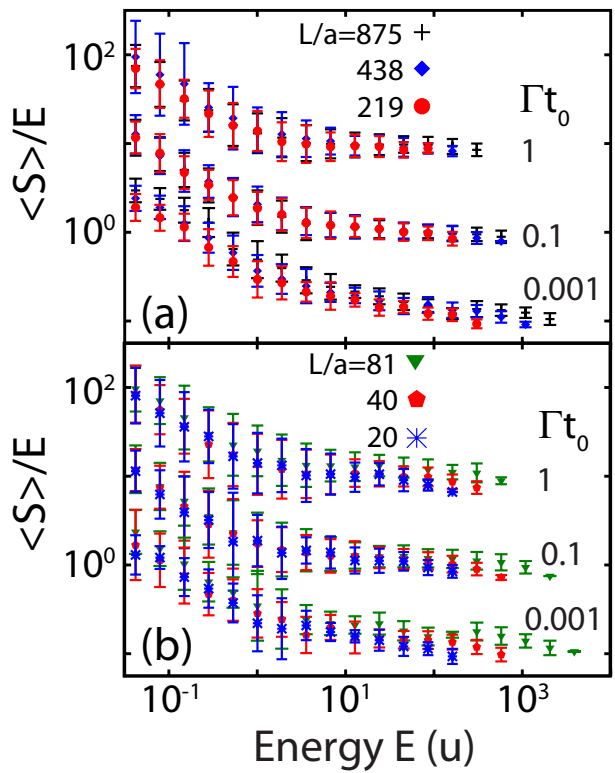}
\caption{\label{S-E-2d-3d} (color online) The ratio of mean stress drop to
energy drop $\langle S\rangle /E$ of the avalanche events for (a) 2D and (b)
3D systems, binned by logarithm of energy.  To prevent overlap, results for
overdamped and underdamped systems are multiplied by 10 and 0.1,
respectively.  Error bars indicate the spread in stress drop for avalanches
of a given energy. A linear relationship between $\langle S\rangle $ and E
holds for $E$ greater than a few $u$ for the overdamped and crossover damping.  }
\end{figure}

Figure \ref{S-E-2d-3d} shows how the mean and variation in $S$ for events
of a given $E$ change with avalanche size.  Results are normalized by $E$
to accentuate deviations from linear behavior and results from different
damping rates are offset to prevent overlap.  For energies less than a
crossover energy the stress drop is much larger than $E$ and has large
fluctuations.  In the overdamped regime this crossover occurs between
$1$ and $2u$ for both 2D and 3D, while for the crossover damping regime we
estimate the crossover energy to occur between $2$ and $4u.$ The presence of
large fluctuations and occasional negative drops suggests that events
smaller than these crossovers do not necessarily contribute to a release of
the imposed shear stress.  For energies larger than the crossover energy
the mean stress drop is nearly equal to E for the overdamped and crossover
damping cases.  Only these larger events exhibit critical scaling for both
$E$ and $S.$ Fig. \ref{S-E-2d-3d} implies that $E$ and $S$ should have the
same scaling exponents in this regime.

In the underdamped limit, $\langle S\rangle /E$ only approaches unity for
the largest events, which move to larger $E$ as $L$ increases.  The sum
rule is not violated, but the scaling of avalanches with $E$ and $S$ may be
different.  The data can be fit to a power law $\langle S\rangle  \sim
E^\eta$ with $\eta \approx 0.9$ over the range $5u < E < 2000u$, but the
prefactor must be $L$ dependent so that $S/E \rightarrow 1$ at the largest
events.  The deviation from linearity is a natural result of reduced
damping and inertia.  In the overdamped limit there should be a
correspondence between stress and potential energy, as traversing each
potential energy barrier dissipates energy.  In the underdamped limit
potential energy barriers may be surmounted with little energy dissipation,
leading to decoupling of the dissipated energy and the stress drop.  The
implications of this decoupling are discussed further in the section on
finite-size scaling.

\subsection{\label{Avalanche Distributions}Avalanche Distributions}

\begin{figure}[placement ]
\includegraphics[width = 0.45\textwidth]
{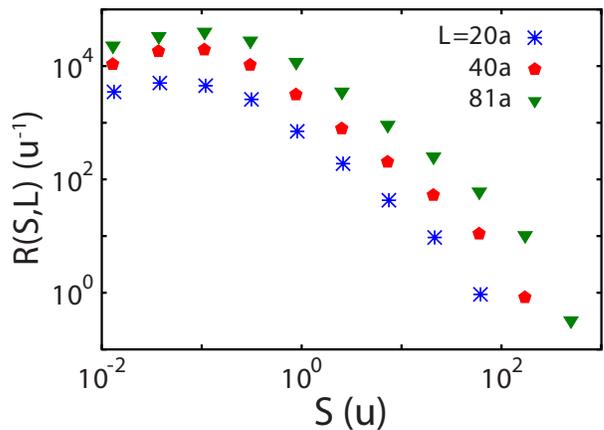}
\caption{\label{raw-dist}(color online) Unscaled distribution of stress
drops $R(S,L)$ for overdamped 3D systems. } \end{figure}

To characterize the different universality classes associated with the
three damping regimes we examine the behavior of the avalanche rate
distribution.  To form this distribution we count the number of avalanche
events with energy drop $E$ or stress drop $S$ during a given strain
interval.  We define the rate of events as the number of events per unit
strain and energy. The rates of events with size $E$ and $S$ in a system of
length
$L$ are denoted $R(E,L)$ and $R(S,L),$ respectively.  

Raw $R(S,L)$ distributions for the overdamped regime in three dimensions
are shown in Fig. \ref{raw-dist}.  As expected, the number of events of a
given size increases with system size.  
One might expect the rate of small avalanches $R(S,L)$ to scale with the number
of particles, \emph{ie} as $L^d.$  This would be the
case if the density of nucleation sites were independent of system size.
Many previous studies of avalanche behavior, for example in interface
depinning, have found or assumed this extensive scaling
\cite{Fisher1998,Martys1991PRB,Stauffer1994}.  In contrast, we find
subextensive scaling in the avalanche rate distribution for all damping
rates in both 2D and 3D.

\begin{figure}[placement h!] 
\includegraphics[width = 0.45\textwidth]
{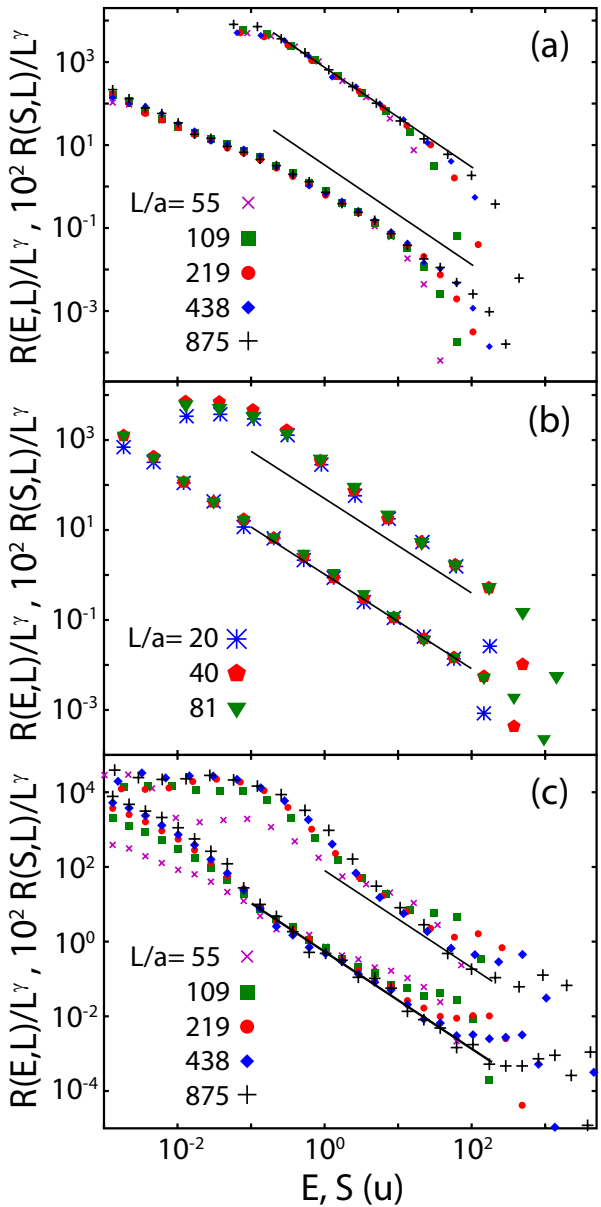}
\caption{\label{gammafull}(color online) Avalanche distributions $R(S,L)$ and $R(E,L)$ 
scaled by $L^{\gamma}$ for the (a) 2D overdamped ($\Gamma t_0 = 1$), (b) 3D
crossover ($\Gamma t_0 = 0.1$) and (c) 2D underdamped ($\Gamma t_0 =
0.001$) regimes.  Values of $\tau$ are given in Table \ref{exponent_table}  }
\end{figure}

Figure \ref{gammafull} shows $R(S,L)$  and $R(E,L)$ scaled by $L^{\gamma}$
with $\gamma$ chosen to collapse the distributions for $E$ and $S$ within
the critical scaling range.  Results for 2D and 3D are similar and only one
example is shown for each $\Gamma.$  For different geometries and damping
rates both $R(S,L)$ and $R(E,L)$ follow a power-law dependence on avalanche
size from $\sim u$ up to a maximum size that grows with system size.  Note
that the nature of the cutoff at large avalanche sizes varies with damping
rate.  There is a simple rapid decay in the number of large events for the
overdamped and crossover cases.  For the underdamped case there is an
excess of large events that leads to a plateau before the distribution cuts
off.

We have already shown in Fig. \ref{S-E-2d-3d} that $E$ and $S$ become
decorrelated for energy drops smaller than a crossover energy.
The distributions $R(S,L)$ and $R(E,L)$ also differ below this scale and only
follow critical scaling for larger events.  For underdamped systems,
$R(E,L)$ and $R(S,L)$ both show $L$ dependent saturation below $E \sim
0.3u$ and $S \sim 2u$, respectively.  For overdamped and crossover systems
$R(S,L)$ saturates for $S \lesssim 0.1$ while $R(E,L)$ continues to rise as
a power law as $E$ decreases.  At the crossover damping, $R(E,L)$ follows a
single power law up to the size dependent cutoff.  For overdamped systems
there is a change in power law at $E \sim u.$ The exponent for small
avalanches is less than unity and varies with system size.  Previous
simulations \cite{Tewari1999,Maloney2004a,Durian1997}
have also observed this regime, but were too small ($L \lesssim
50a$) to see the critical scaling at large E.  Note that $L=55a$ results
are cut off by system size at $E \gtrsim 8u,$ giving less than a decade of
scaling.

Table \ref{exponent_table} lists the values of $\gamma$ that give the best
collapse of $R(E,L)$ and $R(S,L)$ in the critical scaling region from the
crossover energy to the upper cutoff.  Quoted uncertainties indicate where
deviations between curves for different $L$ differ by more than the
statistical errors, which are comparable to the symbol size.  
As noted above, $\gamma$ is substantially less than $d$ in all cases.  This
represents a breakdown of hyperscaling. One possibility is that the
same local nucleation site is likely to produce a bigger avalanche
in larger systems, because there are more surrounding regions to trigger.
Another is tied to the growth in the size of the largest events with $L$.
These larger avalanches may reduce
the probability that a given region can nucleate small events.
The size of the largest avalanches increases with decreasing $\Gamma$ and
there is a corresponding drop in $\gamma.$  

Note that for the overdamped regime the stress distribution shows a larger
range of power-law scaling in Fig. \ref{gammafull}, while in the crossover
and underdamped regimes the energy drop shows a larger range of power-law
behavior.  The deviation in behavior of $S$ and $E$ comes from the regions
where $\langle S \rangle/E > 1 $ in Fig. \ref{S-E-2d-3d}.  This region
extends to larger $E$ as $L$ increases for underdamped systems.  There is a
corresponding shift to larger $S$ in the start of the scaling regime in
$R(S,L)/L^{\gamma}.$

Given the above observations, the most accurate exponents are obtained from
$R(S,L)$ in overdamped systems and $R(E,L)$ for crossover and underdamped
systems.  The difference is only significant for the underdamped case.  In
fact, the value of $\gamma$ recorded for the underdamped regime is
substantially different from the previously reported value of $1.2$ in Ref.
\cite{Salerno2012}.  This is because previous analysis had regarded the
collapse of $R(S,L)$ with more importance and had considered small events
($S<u$) in the fit.  The current collapse puts more emphasis on the quantity
$R(E,L)$ and only attempts to collapse the scaling region $E > u.$ 

The solid lines in Fig. \ref{gammafull} show power-law fits $R(\chi,L) \sim
\chi^{-\tau}$ with $\tau$ given in Table \ref{exponent_table} and $\chi=S$
for overdamped systems and $\chi=E$ for other cases.  Parallel lines are
drawn near $R(E,L)$ for overdamped systems and $R(S,L)$ for other cases.
The results are clearly consistent with power-law scaling over three
decades in event size, but the precise limits of the critical region where
the slope of the distributions should be fit is difficult to determine
using this figure.  As in other critical systems, finite-size scaling of
results for different $L$ provides a better method for determining the
range of critical scaling for the avalanche rate distribution
\cite{privman1990finite,Dahmen1999}.

\subsection{\label{FSS}Finite-Size Scaling}

The assumption underlying the finite-size scaling procedure is that rather
than depending separately on $S$ or $E$ and $L$, the avalanche rate
distributions are a function only of the ratio of avalanche size to a power
of the system size \cite{privman1990finite}. They then obey the scaling ansatz
\begin{equation} R(\chi,L) = L^{\beta}g(\chi/L^{\alpha}), \label{eq:ansatz}
\end{equation} where $\chi$ is either $E$ or $S$ and $g(x)$ is a scaling
function that depends on damping rate $\Gamma$ and may be different for $E$
and $S$.  The scaling function decays to zero at large arguments so that
there are few avalanches above a largest size $\chi_{max}$ that increases
with system size as $L^{\alpha}.$ Given the assumption that no smaller
energy or length scales are important, $g(x)$ must scale as a power law at
small arguments: \begin{equation} g(x) \sim x^{-\tau} \quad , \quad  x <<
1.  \label{scaling-function}\end{equation}

As shown above, the number of avalanches of a given size $\chi$ scales as
$L^{\gamma}$ for $\chi < \chi_{max}$.  
Combining equations \ref{eq:ansatz} and \ref{scaling-function} gives 
\begin{equation}R(\chi,L) =
L^{\beta}g(\chi/L^{\alpha}) \sim
L^{\beta+\alpha\tau}\chi^{-\tau}.
\end{equation} 
This gives us our first scaling relation between exponents, 
\begin{equation}\gamma =
\beta+\alpha\tau. \label{eq:gamma_scaling}\end{equation}

Another scaling relation can be derived from energy balance in steady
state.  The total work per unit volume per unit strain is just the mean
stress $\langle \sigma_s\rangle.$  Equating the total work done in the
entire system to the sum of energy drops in all avalanches one finds:
\begin{equation} \int R(E,L)E dE = \langle \sigma_s\rangle
L^d.\end{equation} 
Inserting the scaling relation and changing variables to
$x=E/L^{\alpha},$ one finds: \begin{equation}  L^{\beta+2\alpha} \int g(x)x
dx \sim L^d, \label{energy_integral}\end{equation} yielding
\begin{equation}\label{eq:bedm2a} \beta = d-2\alpha.
\label{eq:scaling}\end{equation} Note that the integral in Eq.
\ref{energy_integral} converges and is insensitive to the lower bound
because $\tau < 2$ for all systems.  If hyperscaling was obeyed, $\gamma =
d$ would imply $\tau = 2,$ which is clearly inconsistent with the data.

\begin{table} \begin{tabular}
{| c | c | c | c | c | c |} 
\hline $\Gamma$ & d & $\tau$ & $\alpha$ &  $\gamma $ & $\phi$ \cr \hline 
1.0 & 2 & $1.3 \pm 0.1 $ & $0.9 \pm 0.05 $ 
& $1.3 \pm 0.1 $ & $1.00 \pm 0.1 $\\ 
0.1 & 2 & $1.0 \pm 0.05 $ & $0.8 \pm 0.1 $ 
&  $1.2 \pm 0.1 $ & $0.9  \pm 0.1 $\\ 
0.001 & 2 & $1.25 \pm 0.1 $ & $1.6 \pm 0.1 $ 
& $0.8 \pm 0.1 $ & $0.5 \pm 0.1 $\\
\hline 
1.0 & 3 & $1.3 \pm 0.1  $ & $1.1 \pm 0.1 $ 
& $2.1 \pm 0.1 $ & $1.5 \pm 0.2 $ \\ 
0.1 & 3 & $1.05 \pm 0.05 $ & $1.5 \pm 0.1 $ 
& $1.6 \pm 0.1 $ & $ 1.30 \pm 0.1 $\\ 
0.001 & 3 & $1.2 \pm 0.1 $ & $2.1 \pm 0.2 $ 
& $1.3 \pm 0.2 $ & $ 0.9 \pm 0.1 $\\ 
\hline 
\end{tabular}
\caption
{\label{exponent_table}
 Scaling exponents determined for overdamped ($\Gamma t_0=1$) and
underdamped ($\Gamma t_0=0.001$) limits and at the crossover
regime $\Gamma t_0=0.1$ in two and three dimensions.
 Quoted values
satisfy the scaling relation $\gamma=d-(2-\tau)\alpha$ and errorbars are
estimated from the finite-size scaling collapses for $E$ and $S$.
The probability of avalanches decays as $E^{-\tau}$, the largest
avalanche scales as $L^\alpha$, the rate of small avalanches
scales as $L^\gamma$, and the range of stresses scales as $L^{-\phi}$. }
\end{table}

 \begin{figure*}[placement ] \includegraphics[width =
1.0\textwidth]{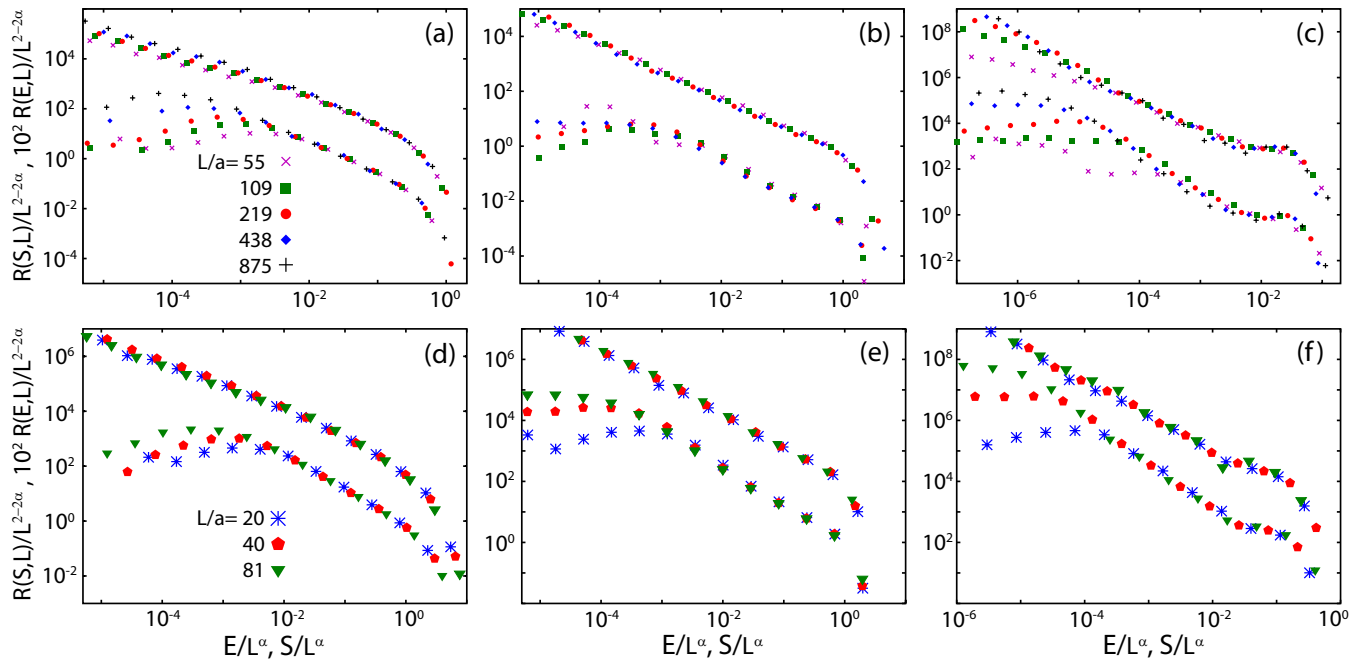} \caption{\label{fssfig} (color
online) Finite-size scaling collapse for $R(E,L)$ and $R(S,L)$
distributions in two dimensions (top) and three dimensions (bottom) for
(a) and (d) overdamped, (b) and (e) crossover, and (c) and (f)
underdamped regimes.  The value
of $\alpha$ used in each collapse is given in Table \ref{exponent_table}
and symbol sizes are comparable to statistical errorbars. } \end{figure*}

Figure \ref{fssfig} shows finite-size scaling collapses for both the energy
and stress drop using the scaling ansatz in Eq. \ref{eq:ansatz}, with
$\beta$ obeying Eq. \ref{eq:bedm2a}.  The exponent $\alpha$ is chosen so
that data for large events from different system sizes collapse onto a
universal curve that corresponds to the scaling function 
$g(\chi/L^{\alpha}).$  
In all cases the curves deviate from the scaling function at a scaled
energy $E \sim u/L^{\alpha},$ which decreases with increasing $L.$  The
energy of a single bond ($\sim u$) is a natural discrete energy scale where the
assumption of scale invariance underlying Eq. \ref {eq:ansatz} breaks down.
We also considered collapses where $\beta$ was allowed to deviate from the
scaling relation in Eq. \ref{eq:bedm2a}, but found there was no significant
improvement.  As with the exponent $\gamma,$ the uncertainties in the value
of $\alpha$ are determined by varying $\alpha$ and  finding a range of
values over which the collapse is acceptable.  This determination utilizes
the fact that the symbol sizes in Fig. \ref{fssfig} are comparable to the
errorbars.

The exponent $\tau$ listed in Table \ref{exponent_table} is found by
multiplying the distributions $R(\chi,L)$ by $\chi^{\tau}$ in order to make
the curves flat over the range of energies from $u$ to $\chi_{max}$
(figures not shown).  The uncertainty in the exponent $\tau$ is determined
from the range of values over which the distributions appear approximately
flat.  The values of the exponent $\tau$ given in Table
\ref{exponent_table} are consistent with the scaling relation Eq.
\ref{eq:gamma_scaling} and fit the unscaled data shown in Fig.
\ref{gammafull}. 
In the overdamped regime fits to $R(S,L)$ give the smallest uncertainty and
in the other cases fits to $R(E,L)$ extend over the longest range.  Fits to
$S$ and $E$ only differ significantly for the underdamped case.
The slope of $R(E,L)$ is given in Table \ref{exponent_table}, while the
slope of $R(S,L)$ appears larger for both 2D and 3D, closer to $\tau
\approx 1.5$ for certain energy ranges.  This difference in slope can be
explained by the features in Fig. \ref{S-E-2d-3d}.  Since $S$ and $E$ are
not linearly related, their distributions should also differ slightly, with
$R(S,L)$ being steeper.  As with the exponent $\gamma,$ our previous paper
\cite{Salerno2012} reported a larger value for $\tau.$  This steeper slope
reflects a value consistent with the distribution $R(S,L)$ as opposed to
$R(E,L).$

It is clear from the finite-size scaling collapses that dimensionality
does not affect the function $g(x),$ but that its form changes with damping
rate.  The form of $g(x)$ in the underdamped regime is of particular
interest.  It displays a characteristic plateau at large avalanche sizes in
both two and three dimensions.  Such an excess of large avalanches is seen
in both earthquakes and experiments on sand \cite
{scholz2002mechanics,Held1990,Baldassarri2006}.  An excess of system spanning events has
also been seen in the Burridge-Knopoff model.  In some versions of that
model a consistent finite-size scaling collapse was not found because a
high-energy peak separated from the lower part of the distribution
\cite{Carlson1989,Carlson1991a,Carlson1991}.  In our system there is a
plateau rather than a second peak, and the entire distribution collapses at
large scaled energies.

\subsection{\label{Plastic}Spatial Extent of Avalanches}

The goal of this section is to relate the spatial extent of the plastic
damage produced by avalanches to the corresponding energy and stress drops.
This is complicated by the long range of elastic interactions.  The
simplest type of shear displacement involving a local rearrangement of a
few atoms produces elastic strains that decay as $r^{-d}$ where $r$ is the
distance from the atoms and $d$ the dimension
\cite{Caroli2009,Maloney2006a,picard2004,Tanguy2006}.  A threshold must be
introduced to distinguish these small elastic strains from the plastic
deformations in the central region.  Appendix B describes how deviations
from the power law decay of strain fields can be used to determine the
threshold used in this section.  Note that this threshold changes
quantitative prefactors in the following discussion, but does not affect
any of the general conclusions.

To define strain fields we first find the displacement of each atom during
an avalanche.  Previous work has emphasized the importance of subtracting
any affine component of these displacements that reflects deformation of
the box \cite{Tanguy2006,Maloney2004,Lemaitre2007}, but this contribution
vanishes in our simulation because no strain is imposed during the
avalanche.  The derivative of the displacement field is calculated by
taking a finite difference of displacements on nearby atoms.  In two
dimensions we form a Delaunay triangulation of the particle positions.
A linear fit to the displacements of the particles on the corners of each
triangle gives $\partial u_i/\partial x_j,$ the derivative of the
displacement $\mathbf{u}$ along direction $i$ with respect to $x_j$
\cite{Maloney2008}.  The
symmetrized strain tensor $\epsilon_{ij}=1/2(\partial u_i/\partial
x_j+\partial u_j/\partial x_i)$ is then constructed to eliminate the effect
of any translation or rotation of the triangle.  In three dimensions, the
strain is obtained from finite differences on a tetrahedral tiling.

\begin{figure}[placement h!]
 \includegraphics[width=0.400\textwidth]
{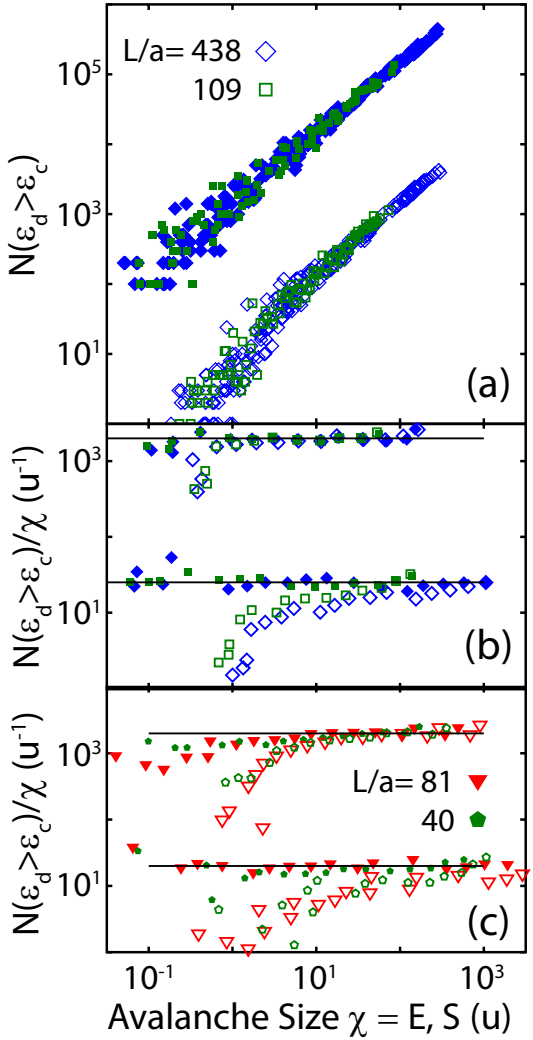}
\caption{\label{Ngt2D}(color online) (a) The number of plastically deformed
Delaunay triangles ( $\epsilon_d > 0.22$) during a plastic event
versus stress drop $S$ (open symbols) and energy drop $E$ (closed symbols)
for two dimensional systems of the indicated size at crossover damping.  Data
for $E$ has been multiplied by 100 to prevent overlap.  (b) Ratio of number
of triangles to mean event size $\chi$ with
$\chi = E$ (closed symbols) and $\chi = S$ (open symbols) for $\Gamma t_0=
0.001$ and $\Gamma t_0 = 1$. 
(c) Ratio of number of plastically deformed tetrahedra to $\chi$ for 3D
systems of the indicated size at $\Gamma t_0= 0.001$ and $\Gamma t_0 = 1$.
Results for $\Gamma t_0 = 1$ in (b) and (c) are multiplied by 100 to
prevent overlap.
} \end{figure}

The magnitude of the strain is usually quantified by rotational invariants.
The first, the trace of the strain tensor, measures the magnitude of
dilational strains.  Shear is most simply related to the second deviatoric
strain invariant $J_2$.  
We define \begin{equation} 
\epsilon_d \equiv \sqrt{J_2} = \sqrt{1/2Tr(\stackrel{\leftrightarrow}{\epsilon}_{dev}^2)}, 
\end{equation} where $\stackrel{\leftrightarrow}{\epsilon}_{dev} $ 
is the deviatoric strain tensor $\stackrel{\leftrightarrow}{\epsilon}_{dev}
\equiv \stackrel{\leftrightarrow}{\epsilon}-
d^{-1}Tr(\stackrel{\leftrightarrow}{\epsilon})
\stackrel{\leftrightarrow}{I}.$ 
In the case of a simple shear strain $\epsilon_s$ in the x-y plane, then
$\epsilon_s = \epsilon_d.$ Triangles or tetrahedra with $\epsilon_d$
greater than a threshold value $\epsilon_c$ are identified as plastic.
Based on the results of Appendix B, we use $\epsilon_c=0.22$ in both 2D and
3D. This is comparable to the ideal elastic limit in dislocation-free
crystals.  Similar results are obtained with other thresholds and by using
the dilational strain.

Figure \ref{Ngt2D}a shows plots of the number of plastically deformed
triangles $N(\epsilon_d > \epsilon_c)$ vs. event size 
for $\Gamma t_0 = 0.1.$  The data for
energy drop $E$ (open symbols) have been multiplied by 100.  Events in the
scaling range ($S,E > u$) show a linear relation between the event size ($S$
or $E)$ and area of the plastic deformation.  Data for $\Gamma t_0 = 1$ and
$\Gamma t_0 = 0.001$ are similar.  

It is not obvious that the spatial extent and energy of events must be
proportional.  In particular, larger events could be associated with
greater dissipation in each spatial region rather than a spread to new
regions.  To test this we found the average $E$ or $S$ of events with a
given spatial size.  Fig. \ref{Ngt2D}b shows the ratio of spatial size to
mean energy in the overdamped and underdamped limits. Results for different
$\Gamma$ are offset to avoid overlap. For the overdamped data the spatial
size of systems is proportional to both $E$ and $S$ for events in the
scaling region identified in previous sections ($S>2u$ and $E>0.3u$). The
energy and spatial size are also proportional for the underdamped case. In
contrast, results for $S$ only asymptote to a linear relation for the
largest events, which grow in size as $L$ increases. This deviation is
further evidence that $E$ is the most natural quantity for the finite-size
scaling collapses of underdamped systems.

The straight lines drawn in Fig \ref{Ngt2D}b are the best fit for the
number of plastically deformed triangles per unit energy.  The values are
about $20u^{-1}$ for the underdamped systems and $18u^{-1}$ for the
overdamped systems.  The constant energy dissipation per unit area is
consistent with limited local plasticity and local particle displacements
during avalanche events.  Such behavior was found in previous
2D simulations, where the total
non-affine displacement of particles over strain
intervals of $1/L$ was at most $a$ \cite{Maloney2008}.
Displacements by a single
particle diameter are sufficient to completely change the local forces and
thus the shear stress driving further deformation.

Fig. \ref{Ngt2D}c shows that the plastically deformed volume also scales
linearly with event energy in three dimensions.  As in 2D, the stress drop
in underdamped systems is less simply related to the plastic volume.  The
horizontal lines in Fig. \ref{Ngt2D}(c) indicate that the number of plastic
tetrahedra per unit dissipated energy is about $20u^{-1}$ for all
damping rates.  This result and the corresponding value for 2D explain the
limit of the scaling region to energies of order $0.1u$ and above.  At
$0.1u$ there are only a handful of triangles or tetrahedra that deform
plastically. It is natural that the finite size scaling ansatz breaks down
and the discreteness of the system becomes important when events involve
only a few particles.

\begin{figure}[placement h!]
 \includegraphics[width=0.400\textwidth]
{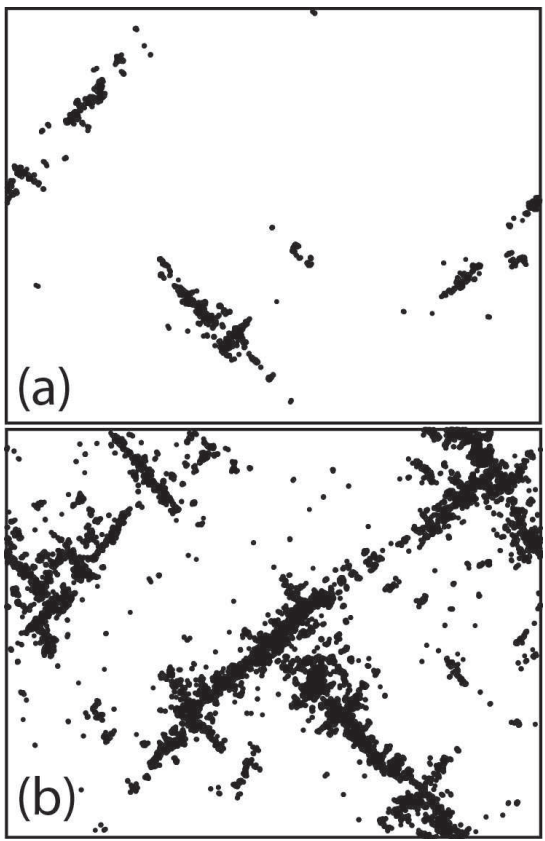}
\caption{Spatial distribution of plastically deformed
triangles in examples of the largest events in
(a) overdamped and (b) underdamped systems with $L=875a$.
The overdamped avalanches have gaps that are large compared to the plastic regions, while the overdamped avalanche spans the system with only small gaps.
Both show strong anisotropy.
\label{realspace}
} \end{figure}

Given the linear relation between energy and the total size of the plastically
deformed region, the largest events involve $\sim L^\alpha$
triangles or tetrahedra.
This implies that $\alpha$ is an effective fractal dimension.
Typical examples of large avalanches are shown in Fig. \ref{realspace}.
In the overdamped limit avalanches contain a number of disconnected
regions that tend to lie along diagonal lines.
While the disconnected regions span the system, they are separated by
larger gaps and 
thus the fractal dimension is less than unity.
In the underdamped limit, $\alpha > d-1$ and the deformed region spreads
across the system with only small breaks.
The clusters are still highly directional with correlations along
the diagonals that have been discussed in past studies of continuous
and lattice models \cite{Maloney2009,Talamali2011}.
These striking changes in avalanche geometry with $\Gamma$
will be discussed in detail in a later
paper and represent another qualitative difference between underdamped
and overdamped systems that may be readily accessible to experiments.

\subsection{\label{Stress}Distribution of Stress Values}

\begin{figure}[placement h!]
\includegraphics[width=0.4\textwidth]
{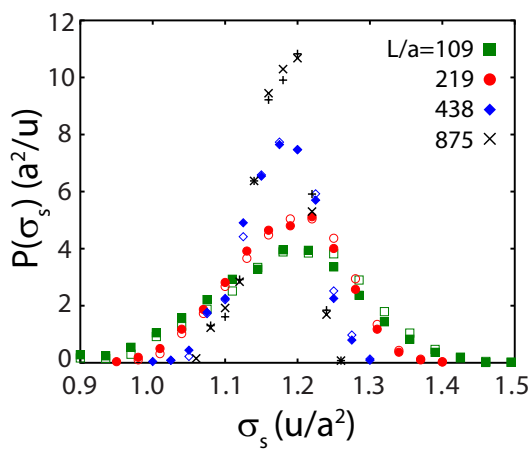}
\caption{\label{raw-stress}(color online) The probability distribution for
stress values $P(\sigma_s)$ before (closed symbols and x) and after (open symbols and +) each avalanche event in the
two-dimensional system in the underdamped regime.  } \end{figure}

One of the most basic quantities measured in a deformation simulation or
experiment is the stress.  In this section we consider the distribution of
shear stress values values before and after each event, $P(\sigma_s).$ 
Figure \ref{raw-stress} shows the distribution of stress values before
(closed symbols) and after (open symbols) avalanches for 2D underdamped
systems of different size.  The distributions narrow about a limiting mean
value as the system size increases.  If inertia drove the system away from
criticality and the onset of shear was a first order transition with
hysteresis, one would
expect a gap between the distribution of stresses before and after
avalanches.  There is no evidence of this separation in our results.  Even
as they narrow, the distribution of stresses before and after avalanches
continue to overlap.  For all cases considered, the shift between the two
distributions is much smaller than their width.  In the following we 
combine the two distributions to improve our statistics.  The distribution
of all instantaneous values of stress gives similar results.

One way to describe the variation in $P(\sigma)$ with system size is to use
a finite-size scaling ansatz similar to Eq. \ref{eq:ansatz} above.   The
shear stress distribution $P(\sigma_s)$ can be rewritten with a scaling
function $h(x)$ as \begin {equation} P(\sigma_s) = L^{\phi}
h(\tilde{\sigma_s}L^{\phi}) \label{eq:stress_ansatz} \end {equation} where
$ \tilde{\sigma_s} = (\sigma_s-\langle \sigma_s\rangle _{L}),$ is the
stress value with the system size dependent mean stress, $\langle
\sigma_s\rangle_L,$ subtracted.  The width of the distribution around the
mean decreases as $L^{-\phi}$ with increasing $L.$ Note that $L$ must enter
with the same power inside and outside the scaling function in order to
preserve the normalization of the probability distribution.

\begin{figure}[placement h!]
\includegraphics[width=0.4\textwidth]
{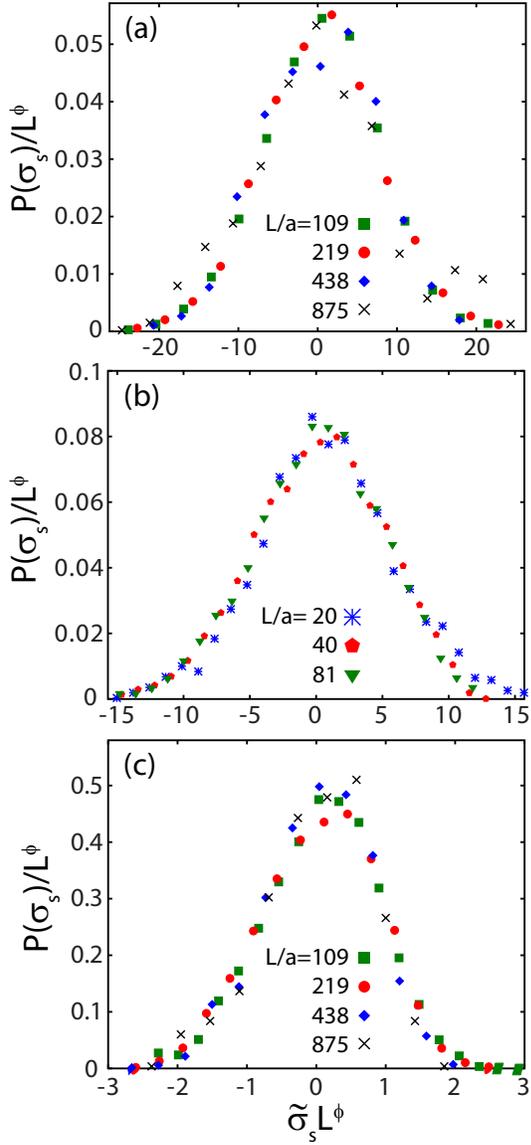}

\caption{\label{scaled_stress}(color online) The scaled distribution of
stress values $P(\sigma_s)$ before and after each avalanche event for the
(a) overdamped system in two dimensions, (b) crossover damping in three
dimensions, and (c) underdamped system in two dimensions.  While the mean value
is approximately constant for each damping rate, the distribution width
scales with system size as $L^{-\phi}.$ Values of $\phi$ are listed in Table
\ref{exponent_table}. Errorbars are comparable to symbol size except for $L/a=875$, where they are up to 5 times larger.
} 
\end{figure}

We confirm the scaling form for the shear stress probability distributions
given in Eq. \ref{eq:stress_ansatz} by finding values of $\phi$ which
collapse results for the various system sizes and damping rates.  Three
such collapses are plotted in Fig. \ref{scaled_stress}, while the best-fit
values of $\phi$ for all systems simulated are listed in Table
\ref{exponent_table}. 

Two upper bounds on the value of $\phi$ may be set.  If there were an
incoherent addition of stress from different regions with no correlations
in time or space, one would expect the width of the distribution to scale
as the inverse square root of the number of independent regions (or
particles).  This yields a relation $ \sqrt{\langle (\sigma_s- \langle
\sigma_s \rangle)^2 \rangle} \sim L^{-d/2},$  or $\phi = d/2.$ Correlations
could make fluctuations decay more slowly with L, so that $d/2$ is an upper
bound for $\phi.$  The width of the distribution must also be at least as
large as the stress change due to the largest avalanches, which are of
order $L^{-(d-\alpha)}.$ This implies that $\phi \leq d-\alpha.$

The values of $\phi$ listed in Table \ref{exponent_table} satisfy the above
bounds in all cases in 2D and 3D, and $\phi$ is comparable to the smallest
bound, $\phi \approx min(d/2,d-\alpha).$
The 2D overdamped and critically
damped systems have $\phi \approx d/2 \approx d-\alpha$.  
In 3D it is clear
that for the overdamped systems $d-\alpha$ is greater than the $d/2$ bound,
and $\phi \approx 1.5 = d/2.$  
Finally, in the underdamped regime $\alpha = 1.6$ in 2D and 
$\alpha = 2.1$ in 3D, and in both cases $\phi \approx d-\alpha.$ 
The crossover from $\alpha < d/2$ to $\alpha > d/2$ is
equivalent to the crossover from positive to negative $\beta,$ indicating
that the rate of events of size $S_{max} \propto L^{\alpha}$ is decreasing,
or the strain interval between such events is increasing.  
It appears that
at this crossover, the events at $L^{\alpha}$ begin to dominate and set the
width of the stress distribution.

The conclusion that the largest events can set the width of the stress
distribution seems inconsistent with Fig. \ref{raw-stress}.  There we found
that the distributions of stresses before events and after events were
nearly the same.  The resolution of this discrepancy is that most events
are small and can occur at any stage of the loading.  The mean and standard
deviation of the stresses before and after small events are
indistinguishable from the global distribution.

For systems with $\alpha < d/2$ even the largest events have a similar
distribution.  This is evident in Fig. \ref{stress-pe-strain} for the
overdamped case where the largest events are smaller than the spread in
stress and occur at all stresses.  For the underdamped case Fig.
\ref{stress-pe-strain} is dominated by the large events which seem to have
a characteristic scale and time interval.  These large events are in the
plateau region where the finite-size of the system is important.  While
they remain the main source of fluctuations in stress for all $L$, the
fractional change in shear stress goes to zero as $L$ increases, because
these largest events increase in size more slowly than $L^d$.

Studies of depinning often control the driving stress rather than the driving
rate \cite{Fisher1998,Martys1991,Friedman2012,Alava2006}.
There is then a critical exponent $\nu$ relating the
distance from the critical stress to the correlation
length $\xi$, corresponding to the linear dimension of the largest
avalanches:
$\xi \sim |\sigma_c-\sigma|^{-\nu}$.
While we have performed simulations with constant rate,
the fact that the range of stresses scales as $L^{-\phi}$ suggests
that $\nu=1/\phi$.
This relation applies in the limit where the largest
avalanches set the range of stress fluctuations, giving
$\nu=1/\phi=1/(d-\alpha)$.
In the case where $\phi=d/2$, stress fluctuations are instead set by
uncorrelated fluctuations in the local properties of the system.
As pointed out by Pazmandi et al. \cite{Pazmandi97},
$1/\phi$ does not correspond to the intrinsic $\nu$ for the correlation
length in this case.

\section{Summary and Discussion} \label{conclusions}

In this paper we have presented a detailed analysis of the dramatic effects
inertia has on quasistatic shear deformation of two and three dimensional
disordered solids.
During the intermittent avalanches of plastic activity, inertia can
carry the system over successive energy barriers and thus
change the ensemble of states sampled.
The most direct evidence for this comes from measurements of the
time dependent
potential energy density and shear stress (i.e. Fig. \ref{stress-pe-strain}).
Damping changes the mean energy density by about 30\% in 2D and 8\% in 3D,
while the shear stress varies by 8\% in 2D and 6\% in 3D.
Even for relatively small systems, $L=200$ in 2D and $40$ in 3D,
there is no overlap between the range of energy densities sampled
in the underdamped and overdamped limits.

Previous studies of lattice models with rules designed to mimic
inertia had predicted profound changes in the nature of the depinning
transition with increasing inertia \cite{Prado1992,Maimon2004,Dahmen2009, Friedman2012,Dahmen2011}.
All found that the onset of shear became hysteretic, with different
stresses needed to initiate and stop motion.
Hysteresis is also present in the recently identified avalanche
oscillator transition \cite{Papanikolaou2012}.
Our simulations of continuous systems show no hysteresis.
In all cases the range of shear stresses and energy densities
sampled during quasistatic
shear goes to zero with increasing system size.
The depinning transition is always continuous and a power law distribution of avalanches
is observed even in the underdamped limit.
However, the scaling exponents are different in underdamped and overdamped limits.

Avalanches were characterized by the total energy dissipated $E$
and an extensive quantity $S$ proportional to the stress drop,
which is more easily measured in experiments.
A sum rule based on conservation of energy requires that the sums over
all avalanches of $E$ and $S$ are equal, but does not relate the two
quantities for individual events.
Indeed, $E$ is always positive, while $S$ can be negative in small systems \cite{Lerner2009}.
At sufficiently high energies, $E$ and $S$ become correlated
and exhibit the same critical scaling.
For high damping this threshold is a fraction of a single particle
bond energy,
but the correlation moves to higher energies in the underdamped limit.
For smaller events,
the distribution of energies (not stresses) can exhibit power law scaling over several decades in $E$
that is not related to the critical behavior of larger avalanches.
Past simulations used smaller system sizes where these noncritical
avalanches dominated the statistics, making it difficult to determine
scaling exponents \cite{Lerner2009,Bailey2007,Maloney2004a}.

Finite-size scaling relations for the rate of avalanches
of a given size in a system of length $L$
were developed and used to determine scaling exponents as a function of damping.
As in our earlier 2D studies, we find three universality classes
corresponding to overdamped and underdamped limits and something
analogous to a multicritical point at a crossover damping that separates them.
Table \ref{exponent_table} summarizes the numerical results for the scaling exponents.

One surprising aspect of the results is that the rate of small events
is not proportional to the system size.
If the probability that a local region would nucleate a small event
was independent of system size, then the rate of small events
would grow as $L^d$.
The observed rate grows as $L^\gamma$ with $\gamma$ significantly less than $d$.
One possibility is that local configurations that would only produce a small
avalanche in a small system can trigger a string of other
instabilities in a larger system.
The long-range power law decay of elastic interactions makes it more
likely that they can affect scaling exponents in this way. 
The size of the largest avalanches grows with system size, and it is also
possible that these largest avalanches lower the number of nucleation
sites for small avalanches.
The difference between $d$ and $\gamma$
for all damping regimes in two and three dimensions
represents a violation of hyperscaling.
To our knowledge, values of $\gamma$ have not been reported for
lattice models, but the restriction to discrete changes on discrete lattice sites may
lead to different scaling.

The scaling exponent $\alpha$ reflects the growth the of the largest
avalanches with system size $E_{max} \propto L^{\alpha}.$  
Values of $\alpha$ in all systems are found to be less than spatial
dimension $d$. In the overdamped limit $\alpha$ is slightly lower than unity in 2D
and slightly larger than unity in 3D.
Previous simulations of smaller overdamped systems
(less than 20,000 particles) reported $\alpha=1$ \cite{Maloney2004a}
or $0.74$ \cite{Lerner2009} in 2D, and about $1.5$ in 3D \cite{Bailey2007}, but with significant uncertainties
and poor collapses at the largest energies.
Lattice models of overdamped avalanches that include the tensoral nature of shear stress find $\alpha=1$ \cite{Zaiser2007,Talamali2011}.
This is quite close to our simulation results, while scalar models predict $\alpha=2$ \cite{Dahmen2009,Fisher1997}.
The large difference in predictions for $\alpha$ make it a useful quantity to measure in future
experiments, but we know of no existing studies.
As noted above, $\alpha$ is effectively a fractal dimension for the plastically deformed region
which is generally disconnected and strongly anisotropic \cite{Maloney2006a,Maloney2009}

The largest avalanches grow much more rapidly with system size in the underdamped limit,
$\alpha=1.6\pm 0.1$ and $2.1\pm 0.2$ in 2D and 3D, respectively.
As a result, these large avalanches dominate the fluctuations in instantaneous
stress values, which scale as $L^{-\phi}$ with $\phi=d-\alpha$.
Plots of the evolution of stress with time are dominated by these large events.
In the overdamped regime, $\phi=d/2$, indicating that fluctuations are dominated
by uncorrelated variations in interactions and geometry as the
configuration of particles evolves.

Studies of avalanche statistics in slowly driven systems have generally
focussed on the exponent $\tau$ that describes the decrease in event
rate with event size: $R(E) \sim E^{-\tau}$.
Our simulations reveal similar values of $\tau \approx 1.25$
for both underdamped and overdamped limits in 2D and 3D.
The crossover damping has a lower value of $\tau \approx 1$.
Direct fits of power laws over at least 3 decades
are consistent with exponents obtained from finite-size scaling
relations.
The only discrepancy is for scaling of stress drops in the underdamped
limit.
In this case,
the stress drop rises less rapidly than the energy up to the size
of the largest avalanches.
This may be because inertia is more likely to carry the system past
barriers that lead to a lowering in stress than energy.
The result is an apparent $\tau$ of about 1.5 over a limited range of stress
drops and a poorer finite-size scaling collapse.
The fact that $S$ and $E$ are different in the underdamped limit may have
implications for experimental studies where stress or slip displacements are often more
directly accessible than energy.

Lattice models of overdamped dynamics that treat strain as a scalar variable predict a universal
value of $\tau=3/2$ that is clearly inconsistent with our scaling relations.
A recent model that includes the directional nature of stress transfer
yields a consistent value of about 1.25 \cite{Talamali2011}.
As noted above, lattice models predict hysteresis in the underdamped limit.
This is inconsistent with our results, but the hysteresis is related to a 
discrete change from inertia and local rearrangements produce discrete
drops in strain.
It would be interesting to explore the effect that continuous changes
from inertia and/or in local stress would have on scaling in lattice models.

A critical review of experimental results for $\tau$ is beyond the scope
of this paper, but the wide range of reported values reflects
systematic uncertainties in measuring avalanches and analyzing their
statistics.
In many cases experiments are not in the steady state regime considered here,
but collect avalanche statistics at small strains during the transition from
the inital elastic response to yield (i.e. $\epsilon < 0.07$ in Fig. \ref{stress_strain}).
We find that this initial region is sensitive to preparation and exclude
it from our analysis.
In most cases the range of avalanche sizes is only one to two decades.
Uncertainties in the limits of the scaling regime due to finite system size,
finite temperature \cite{Hentschel2010},
distance from the critical stress \cite{Friedman2012}
and experimental noise lead
to uncertainties that are greater than the difference
between $\tau=1.5$ and 1.25.
The finite-size scaling method used here eliminates such systematic effects,
but is difficult to replicate in experiments.

Earthquake statistics cover the largest range of event sizes, but involve
a wide range of phenomena that are not included in simple models like ours
\cite{scholz2002mechanics}.
Just one example is the finite thickness ($\sim 10$km) of the active region of the crust.
There is a transition in the effective dimension of earthquake displacements
as the size of the slipping region grows past this scale \cite{Pacheco1992,Shaw2009}.
Seismologists also treat successive displacements along different faults
as separate events, even though there is evidence that they are causally
connected.
Our quasistatic simulations group together all plastic activity resulting
from an initial instability even if the activity occurs in widely
separated regions and is separated by long quiescent periods.
These long quiescent periods do not occur in lattice models where
changes in local strain are discrete.
We are currently evaluating methods of separating our avalanches
into separate events, and initial results suggest the effective value
of $\tau$ could increase in the overdamped limit where quiescent periods
are more common.

Given the difficulties in determining precise values of $\tau$,
it becomes interesting to
consider the qualitative difference in the scaling function $g$ for
overdamped and underdamped systems.
Models and experiments on overdamped systems generally find
a simple exponential cutoff for large events \cite{Maloney2004a,Bailey2007,Lerner2009,Dennin2003,Dennin2002}.
In contrast,
studies of earthquakes statistics for single faults often find
an excess of large system spanning events that is similar to
the plateau seen in our underdamped simulations.
Plateaus of this type are also evident in experimental studies
of steady shear in glass bead packs \cite{Baldassarri2006}.
Future experiments that focus on the form of the scaling function and
the fractal dimension $\alpha$ of the largest avalanches may provide the
most sensitive means of detecting inertial effects and testing
theoretical models.

\acknowledgments{
This work was supported by the National Science Foundation under Grants 
No. DMR-1006805, IGERT 0801471, CMMI-0923018, and OCI-0963185.
MR acknowledges support from the Simons Foundation.}


\bibliography{3dp}

\appendix \section{Sum Rule Derivation} \label{sum_rule} The sum rule in
Eq. \ref{sum_rule_eq} follows from energy conservation.  Each avalanche
event, $i,$ is accompanied by a potential energy density drop $\Delta
\mathcal{U}^{(i)}$ and shear stress drop $\Delta \sigma^{(i)}.$  This
dissipated energy must be balanced by the work done on the system during
segments where the system loads elastically.  For a strain segment $\Delta
\epsilon^{(j)}$ the work done on the system is $ \sigma_s \Delta
\epsilon^{(j)} L^d$. 

The assumption that there is a well defined steady-state mean potential
energy density allows us to equate the sum of the energy drops with the
total work done 
\begin{equation}\label{worksum} 
\sum_i \Delta \mathcal{U}^{(i)}L^d = \sum_j \sigma_s \Delta \epsilon^{(j)}L^d, 
\end{equation}  where the sum on the left is over all energy drops, and the
sum on the right is over all elastic loading segments, which are
equal in number.

As shown in Figs. \ref{stress_strain}, \ref{raw-stress} and
\ref{scaled_stress} there is also a well defined steady-state shear stress. Thus
we can rewrite Eq. \ref{worksum} as \begin{equation} \sum_i \Delta
\mathcal{U}^{(i)}L^d = \langle \sigma_s\rangle \sum_j \Delta \epsilon^{(j)}L^d,
\end{equation} introducing corrections proportional to the square of stress
fluctuations, which go to zero as $L^{-2\phi}$ in the thermodynamic limit
(Table \ref{exponent_table}).  A steady state shear stress also implies
that the stress rises during elastic loading balance the stress drops
during avalanches over long strain intervals. The stress rise over each
elastic interval $j$ can be written as $\mu \Delta \epsilon^{(j)}$ where
$\mu$ is the shear modulus, so that: 
\begin{equation} \sum_i
\Delta\sigma^{(i)}L^d = \sum_j 4\mu \Delta \epsilon^{(j)}L^d.
\end{equation}  Other workers have found that above a length scale much
smaller than our system sizes variations in the modulus $\mu$ between
different elastic segments are small \cite{Tsamados2009}.

In the main text we measure extensive stress and energy drops $ S \equiv
\langle \sigma_s\rangle L^d/(4\mu)$
and $E = \Delta \mathcal{U}L^d$ in order to compare avalanches
across different system sizes.  Combining the relations above with the
definitions of $S$ and $E$ allows us to relate total stress and energy
drops
\begin{equation} \sum_i S^{(i)} = \sum_i E^{(i)} \end{equation} Since the
summations are over the same set of avalanche events this also implies that
the mean values are equal $ \langle S \rangle = \langle E \rangle.$ 

\section{Distinguishing Plastic and Elastic Regions}
\label{plastic-appendix}

As noted in the main text, the strain field around a local local plastic
region decays as a power of the distance $r$ from the
region\cite{Caroli2009,Maloney2006,picard2004,Tanguy2006}.
The prefactor should be proportional to the magnitude of
the plastic rearrangement, which we find scales as the stress drop.  Since
the spatial arrangements of plastic regions can be complicated, we consider
instead the distribution of local strain values $N(\epsilon_d)$.  From the
scaling of the phase space with distance $r$, we have $r^{d-1} dr \sim
N(\epsilon_d)d\epsilon_d.$  Then the distribution of local strains scales
as a power law in two and three dimensions
\begin{equation} N(\epsilon_d)
\sim S\epsilon_d^{-2}.
\end{equation}
The cumulative distribution function
(CDF) $N(\epsilon_d > x)$ of strains larger than $x$ scales as:
\begin{equation} N(\epsilon_d > x) \sim Sx^{-1}.  \label{cumdist}\end{equation} 

\begin{figure} [Placement h!]
\includegraphics[width=0.35\textwidth]
{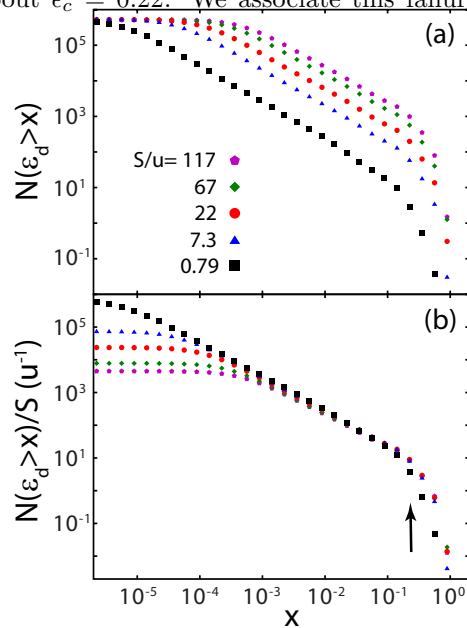}
\caption{\label{2dpofw}(color online) (a) The CDF of
$\epsilon_d$ averaged over events of the indicated $S$ in two dimensions
for damping $\Gamma t_0 = 1$ and system size $L = 438a$.
(b) CDF after rescaling by $S$ to collapse the elastic region.
An arrow indicates $\epsilon_c = 0.22.$} \end{figure}

Figure \ref{2dpofw}(a) shows the $N(\epsilon_d > x)$ averaged
over avalanche events of a given stress drop $S$.  Events with damping rate
$\Gamma t_0 = 1$ are shown, and curves for other damping rates are similar.
Above a minimum strain that grows with $S$, each curve follows the
power law scaling predicted by Eq. \ref{cumdist}.
Eq. \ref{cumdist} also predicts that the prefactor of the power law region
should grow linearly with the size of the event.
Fig. \ref{2dpofw}(b) tests this prediction.
We find that events large enough to be in the scaling regime ($S >u$)
collapse onto a universal curve with a power law regime that
is cutoff at strains bigger than about $\epsilon_c=0.22$.
We associate this failure of the elastic prediction with the onset of
plastic deformation in the main text.
Note that smaller events are cutoff at slightly smaller strains,
providing further evidence that they involve different types of
displacement.

\begin{figure}[Placement h!]
\includegraphics[width=0.4\textwidth]
{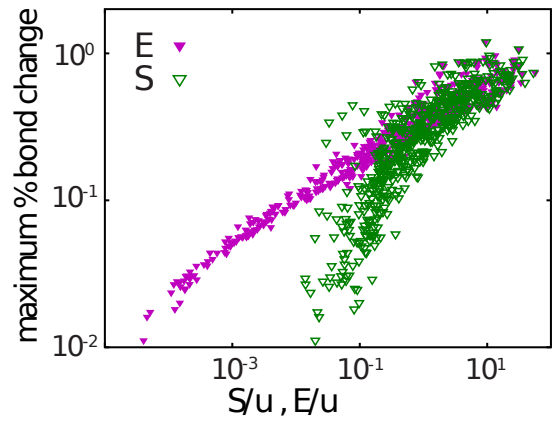}
\caption{\label{bond-change}(color online) The maximum percentage bond
change that occurs during an avalanche of stress drop $S$ (open symbols) or
energy drop $E$ (closed symbols) in overdamped 2D systems with $L=109a.$
Similar results are obtained for other $L$ and $\Gamma.$} \end{figure}

Changes in bond length are another measure of local deformation that can be
used to identify plastic regions. Fig. \ref{bond-change} shows the maximum
percentage change of any bond in the system as a function of event size.
Note that there are almost no events where bonds change less than 2\%.
These are associated with extremely small events of order $E \sim
10^{-5}u.$  For events in the scaling regime where S and E are comparable
($E>0.3u$), the largest bond change is 20\%. This is comparable to the
displacements need to produce a local strain of $\epsilon_c.$

Even the largest events produce only ~100\% bond changes, corresponding to
displacements of order a bond length relative to neighbors. This is
consistent with the conclusion in the main text that larger events produce
a uniform amount of dissipation over larger regions rather than larger
deformations in a fixed spatial region. The very slow increase in the
maximum bond length change  with event size for the largest events may be
attributed to sampling more bond changes from a fixed distribution. This is
consistent with the collapse of the CDF in Fig. \ref{2dpofw}.

\end{document}